\documentclass[twocolumn,aps,notitlepage,showpacs,showkeys]{revtex4}%
\usepackage{amsfonts}
\usepackage{amsmath}
\usepackage{amssymb}
\usepackage{graphicx}
\usepackage{subfigure}%
\begin{document}
\title{Conditional dynamics induced by new configurations for Rydberg dipole-dipole interactions}
\author{E. Brion}
\email{ebrion@phys.au.dk}
\author{A. S. Mouritzen}
\author{K. M\o lmer}
\affiliation{Lundbeck Foundation Theoretical Center for Quantum System Research Department
of Physics and Astronomy, University of Aarhus, Ny Munkegade, Bld. 1520,
DK-8000 \AA rhus C, Denmark}
\date{\today}

\begin{abstract}
We suggest a novel way to use strong Rydberg 
dipole-dipole interactions in order to induce non-trivial conditional dynamics
in individual-atom systems and mesoscopic ensembles.\ Contrary to previous
works, we suggest to excite atoms into different Rydberg states, which results in a
potentially richer dynamical behaviour. Specifically, we
investigate systems of individual hydrogen-like atoms or mesoscopic ensembles
excited into high-lying hydrogen-like $s$, $p$ or $d$ states and show how to perform
three-qubit conditional dynamics on the information they contain through a
proper use of dipole-dipole interaction induced energy shifts.

\end{abstract}
\maketitle

\section{Introduction}
Due to their large dipole moments
\cite{Gallagher}, Rydberg atoms experience strong long-range dipole-dipole
interactions. These interactions strongly mix and shift the multiply Rydberg excited collective states of an atomic sample. This phenomenon has recently been put forward as the key ingredient of different promising atomic
quantum processing scenarios. For instance, Rydberg-Rydberg interactions can be used to perform two-qubit logic operations in individual-atom
systems by shifting a transition off resonance in an atom, depending on the
internal state of another atom in its immediate neighbourhood \cite{Jaksch,
Grangier,Saffman}. In a mesoscopic ensemble, dipole-dipole interactions are
able to inhibit transitions into collective states which contain more than one
Rydberg excitation, thus leading to the so-called Rydberg blockade.\ First
predicted in \cite{Lukin}, this phenomenon was locally observed in laser
cooled atomic systems \cite{Tong, Singer, Cubel,Anderson, Vogt} and could in
principle be used in the future to manipulate and entangle collective
excitation states of mesoscopic ensembles of cold atoms \cite{Lukin}.
\newline\indent So far, the schemes based on Rydberg-Rydberg interactions have
focused on the coupling between atoms excited into the same high energy state.
Typically, in these proposals, atoms in the sample are (simultaneously or not)
excited to the same Rydberg state $\left\vert ns\right\rangle $. When the
so-called F\"{o}rster process $ns+ns\rightarrow np+\left(  n-1\right)  p$ is
resonant, the dipole-dipole interaction is enhanced: one can then diagonalize
the dipole-dipole interaction operator $V_{dd}$ in the subspace $\left\{
\left\vert ns,ns\right\rangle ,\left\vert np,\left(  n-1\right)
p\right\rangle ,\left\vert \left(  n-1\right)  p,np\right\rangle \right\}  $,
which leads to shifted new eigenstates (see \cite{Grangier} for a detailed discussion). In the present paper, we propose to
investigate other settings in which atoms can be excited into several Rydberg states of different $l$'s. In these configurations, $V_{dd}$ mixes and shifts some of the several-atom states through F\"{o}rster-like processes, whereas
it leaves the others unchanged: figuratively speaking, depending on the Rydberg state they are excited into, atoms will
\textit{see} each other or not. This can be used to selectively hinder certain transitions into multiply Rydberg excited states, while allowing for the others. This, in turn, leads to richer dynamical behaviours than considered in previous
theoretical proposals. Through exciting the
information-carrying ground states of the atoms into properly chosen different
Rydberg states, one can for example induce conditional
logic dynamics involving more than two qubits. \newline\indent To be more specific, in the present paper, we
shall first focus on the dipole-dipole interactions which take place in a
system of three hydrogen-like atoms submitted to different laser beams
coupling their ground levels to different Rydberg states $\left\vert
r_{1}\right\rangle $, $\left\vert r_{2}\right\rangle $ and $\left\vert
r_{3}\right\rangle $, respectively. We shall carefully examine the interaction induced energy shifts and the resulting blockade of unwanted transitions, and in particular we shall show that the desired performance is satisfactorily met by rubidium atoms. We shall then demonstrate
how to use this physical setting in order to perform non-trivial conditional
logic operations involving three qubits of information stored either
in a three-atom system or in mesoscopic ensembles, thus generalizing the
pioneering work by Lukin \textit{et al}. \cite{Lukin}. \begin{figure}[ptb]
\begin{center}
\includegraphics[
height=2in,
width=2.3in
]{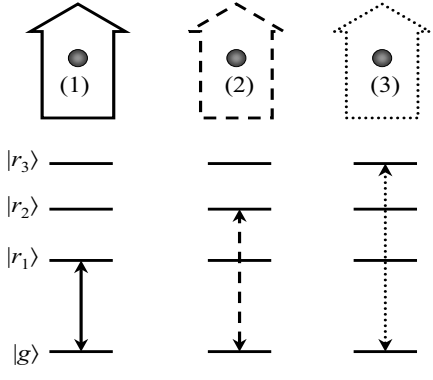}
\end{center}
\caption{Level scheme and laser excitations of the three-atom system.}
\label{fig1}%
\end{figure}
\section{Dipole-dipole interactions between
three hydrogen-like atoms in different Rydberg states}\label{sec2} \indent Let us consider
a system of three identical hydrogen-like atoms, denoted $\left(  1\right)  ,\left(
2\right)  ,\left(  3\right)  $. We let $\widehat{H}_{0,i}$ be the unperturbed
Hamiltonian of atom $\left(  i\right)  $ and
\begin{equation}
\widehat{V}_{i,j}\equiv\frac{1}{4\pi\varepsilon_{0}}\left[  \frac
{\widehat{\overrightarrow{\mu}}_{i}\cdot\widehat{\overrightarrow{\mu}}_{j}%
}{R_{ij}^{3}}-3\frac{\left(  \widehat{\overrightarrow{\mu}}_{i}\cdot
\overrightarrow{R}_{ij}\right)  \left(  \widehat{\overrightarrow{\mu}}%
_{j}\cdot\overrightarrow{R}_{ij}\right)  }{R_{ij}^{5}}\right]  \label{DipDip}%
\end{equation}
be the dipole-dipole interaction between atoms $\left(  i\right)  $ and $\left(
j\right)  $, where $\overrightarrow{R}_{ij}\equiv R_{ij}\overrightarrow
{u}_{ij}=R_{ij}\left(  \sin\alpha_{ij}\cos\beta_{ij}\overrightarrow{e}%
_{x}+\sin\alpha_{ij}\sin\beta_{ij}\overrightarrow{e}_{y}+\cos\alpha
_{ij}\overrightarrow{e}_{z}\right)  $ is the vector from nucleus $\left(
i\right)  $ to nucleus $\left(  j\right)  $. The total
Hamiltonian of the system then takes the form $\widehat{H}=\widehat{H}_{0}%
+\widehat{V}_{dd}$ with $\widehat{H}_{0}\equiv\sum_{i=1}^{3}\widehat{H}_{0,i}%
$\ and $\widehat{V}_{dd}\equiv\sum_{i<j}\widehat{V}_{i,j}$. Moreover, with electronic states
$\psi_{nlm}\left(  r,\theta,\phi\right)  =R_{nl}\left(  r\right)
Y_{lm}\left(  \theta,\phi\right)  $ \cite{BJ03}, one gets the general formula
\begin{align}
& \left\langle n_{i},l_{i},m_{i};n_{j},l_{j},m_{j}\left\vert \widehat{V}_{ij}\right\vert n_{i}^{\prime},l_{i}^{\prime},m_{i}^{\prime};n_{j}^{\prime
},l_{j}^{\prime},m_{j}^{\prime}\right\rangle 
\nonumber \\
& =\frac{e^{2}\mathcal{R}_{n_{i},l_{i}}^{n_{i}^{\prime},l_{i}^{\prime}}\mathcal{R}_{n_{j},l_{j}}^{n_{j}^{\prime},l_{j}^{\prime}}}{4\pi
\varepsilon_{0}R_{ij}^{3}} \nonumber \\
& \times
\left[  \overrightarrow{\mathcal{A}}_{l_{i},m_{i}}^{l_{i}^{\prime}%
,m_{i}^{\prime}}\cdot\overrightarrow{\mathcal{A}}_{l_{j},m_{j}}^{l_{j}%
^{\prime},m_{j}^{\prime}}-3\left(  \overrightarrow{\mathcal{A}}_{l_{i},m_{i}%
}^{l_{i}^{\prime},m_{i}^{\prime}}\cdot\overrightarrow{u}_{ij}\right)  \left(
\overrightarrow{\mathcal{A}}_{l_{j},m_{j}}^{l_{j}^{\prime},m_{j}^{\prime}%
}\cdot\overrightarrow{u}_{ij}\right)  \right] \label{Eq2}
\end{align}
where
\begin{align*}
\mathcal{R}_{n,l}^{n^{\prime},l^{\prime}} &  \equiv\int_{0}^{+\infty} dr \ r^{3}R_{nl}\left(  r\right)  R_{n^{\prime}l^{\prime}}\left(  r\right)  \\
\overrightarrow{\mathcal{A}}_{l,m}^{l^{\prime},m^{\prime}} &  \equiv\int
_{0}^{\pi}d\theta\sin\theta\int_{0}^{2\pi}d\phi\overrightarrow{e}_{r}\left(
\theta,\phi\right)  Y_{lm}^{\ast}\left(  \theta,\phi\right)  Y_{l^{\prime
}m^{\prime}}\left(  \theta,\phi\right)
\end{align*}
and $\overrightarrow{e}_{r}\left(  \theta,\phi\right)  =\sin\theta\cos
\phi\overrightarrow{e}_{x}+\sin\theta\sin\phi\overrightarrow{e}_{y}+\cos
\theta\overrightarrow{e}_{z}$ (see Appendix \ref{appR} for an explicit expression for $\mathcal{R}_{n,l}^{n^{\prime},l^{\prime}}$). Note that $\overrightarrow{\mathcal{A}}_{l,m}^{l^{\prime},m^{\prime}} \neq 0$ only if $l^{\prime} = l \pm 1$ and $m^{\prime} - m = 0, \pm 1$ (dipole selection rules).
 
\indent In our setting, the three atoms, initially prepared in the ground state $\left\vert
g\right\rangle $, can be submitted to different (sets of) laser beams which couple
$\left\vert g\right\rangle $ to the Rydberg states $\left\vert
r_{1}\right\rangle \equiv\left\vert n s\right\rangle $, $\left\vert
r_{2}\right\rangle \equiv\left\vert n p,k_{2}\right\rangle $ and
$\left\vert r_{3}\right\rangle \equiv\left\vert n d,k_{3}\right\rangle $ (see Fig. \ref{fig1}). Population of states with arbitrary magnetic quantum numbers $k_{2}$ and
$k_{3}$ is achieved through a proper choice of the polarization of the
laser beams. If the dipole-dipole interaction were absent, the only
populated three-atom states would be $\left\vert g;g;g\right\rangle $, $\left\vert
r_{1};g;g\right\rangle $, $\left\vert g;r_{2};g\right\rangle $, $\left\vert
g;g;r_{3}\right\rangle $, $\left\vert r_{1};r_{2};g\right\rangle $,
$\left\vert r_{1};g;r_{3}\right\rangle $, $\left\vert g;r_{2};r_{3}%
\right\rangle $ and $\left\vert r_{1};r_{2};r_{3}\right\rangle $. The effect of $V_{dd}$ on the ground state $\left\vert g;g;g\right\rangle $ and the singly Rydberg excited states is very small and we shall neglect it; in contrast, $V_{dd}$ strongly couples the doubly and triply Rydberg excited states to the rest of the Hilbert space. 

Nevertheless, choosing $n$ in such a way that all the couplings listed in Table \ref{TabRub} are non-resonant, we shall assume that we can restrict ourselves to the two resonant couplings $\left\vert ns ; np\right\rangle \leftrightarrow \left\vert np ; ns\right\rangle $ and $\left\vert np ; nd\right\rangle \leftrightarrow \left\vert nd ; np\right\rangle $. Note that, by virtue of selection rules, the states $\left \vert ns, nd \right \rangle$ and $\left \vert nd, ns \right \rangle$, though resonant, are not coupled by $V_{dd}$. The applicability of the previous assumption will be discussed below and quantitative conditions for its validity will be identified. For now, let us assume these conditions are met: the state $\left\vert r_1;g;r_3 \right\rangle$ is then unaffected by $V_{dd}$ (in first order), whereas $\left\vert r_1;r_2;g \right\rangle$, $\left\vert g;r_2;r_3 \right\rangle$ and $\left\vert r_1;r_2;r_3 \right\rangle$ are shifted. The (first order) shifts can be calculated by diagonalizing $V_{dd}$ in the three degenerate subspaces 
\begin{eqnarray*}
\mathcal{H}_{sp} & = & Span\left\{  \left\vert n s;n p;g\right\rangle ,\left\vert n p;n s;g\right\rangle \right\}  ,\\
\mathcal{H}_{pd}  & = & Span\left\{  \left\vert g;n p;n d\right\rangle ,\left\vert g;n d;n p\right\rangle \right\}  ,\\
\mathcal{H}_{spd} & = & Span \left\{
\begin{array}
[c]{c}%
\left\vert ns;np;nd\right\rangle ,\left\vert np;nd;ns\right\rangle ,\left\vert
nd;ns;np\right\rangle ,\\
\left\vert ns;nd;np\right\rangle ,\left\vert nd;np;ns\right\rangle ,\left\vert
np;ns;nd\right\rangle
\end{array}
\right\},
\end{eqnarray*}
where the magnetic quantum numbers $m_p = -1,0,1$ and $m_d = -2,-1,0,1,2$ are implicit.
\newline Using Eq.(\ref{Eq2}) and taking the selection rules into account, one derives the following expression for $V_{dd}$ in $\mathcal{H}_{sp}$:
\[
V_{sp} = \frac{e^2 \left( \mathcal{R}_{ns}^{np} \right)^2 }{4 \pi \epsilon_{0} R_{12}^3} \times A_{sp}, \quad
A_{sp} = \left(
\begin{array}
[c]{cc}
0 & M_{sp}^{\dagger}\\
M_{sp} & 0
\end{array}
\right) ,
\]
where $M_{sp}$ is a $3\times 3$ matrix which contains coupling terms between $\left \vert ns ; np, m_p \right \rangle $ and $\left \vert  np, m_p^{\prime};ns \right \rangle $ of the form $\left[  \overrightarrow{\mathcal{A}}_{s}^{p,m_{p}^{\prime}}\cdot\overrightarrow{\mathcal{A}}_{p,m_{p}}^{s}-3\left(  \overrightarrow{\mathcal{A}}_{s}^{p,m_{p}^{\prime}}\cdot\overrightarrow{u}_{12}\right)  \left(
\overrightarrow{\mathcal{A}}_{p,m_{p}}^{s}\cdot\overrightarrow{u}_{12}\right)  \right]$. $A_{sp}$ has six non-zero eigenvalues $\left\{ \pm \frac{1}{3}, \pm \frac{1}{3}, \pm \frac{2}{3} \right\}$ which do not depend on the geometric configuration of the system (\textit{i.e.} the angles and distances between the atoms). The corresponding energy shifts are given by $\Delta_{sp} = \frac{e^2 \left( \mathcal{R}_{ns}^{np} \right)^2 }{4 \pi \epsilon_{0} R_{12}^3} \times \left\{ \pm \frac{1}{3}, \pm \frac{2}{3} \right\}$. 
\newline Similar results can be established in $\mathcal{H}_{pd}$: the energy shifts one obtains are non-zero and depend neither on distances nor on angles between atoms; their norms take the values in the range $\left\vert \Delta_{pd} \right \vert = \frac{e^2 \left( \mathcal{R}_{np}^{nd} \right)^2 }{4 \pi \epsilon_{0} R_{23}^3} \times \left\{ 0.023 - 0.643 \right\}$.
\newline The case of $\mathcal{H}_{spd}$ is more complicated, since $V_{dd}$ now involves both $sp-ps$ and $pd-dp$ couplings: $V_{spd}$ thus cannot be decomposed in as simple a way as $V_{sp}$ and $V_{pd}$. The resulting eigenvalues will thus depend on the geometry of the three-atom system (one angle- and two distance-variables, for instance). In principle, it is thus possible to find a specific arrangement so that one or more eigenvalues are null; in the generic situation, however, all the eigenvalues and the associated energy shifts $\Delta_{spd}$ are non-zero.
\newline Finally, $\Delta_{sp},\Delta_{pd},\Delta_{spd}$ prevent the states $\left\vert r_1;r_2;g \right\rangle$, $\left\vert g;r_2;r_3 \right\rangle$ and $\left\vert r_1;r_2;r_3 \right\rangle$ from being populated through resonant laser excitation of $\left\vert g;g;g \right\rangle$, whereas all the other unshifted states, and in particular $\left\vert r_1;g;r_3 \right\rangle$, are accessible. 
\newline\indent Before addressing the physical implementation of this situation in a rubidium atom system, let us turn back to the assumptions which allowed us to restrict ourselves to $\mathcal{H}_{sp}, \mathcal{H}_{pd}, \mathcal{H}_{spd}$. These assumptions are legitimate when the second-order shifts induced by the non-resonant couplings shown in Table \ref{TabRub} are negligible compared to the first-order shifts obtained above. To make sure this is fulfilled, one has to verify that the smallest of the first-order shifts, within each subspace, $min \left( \Delta^{(1)} \right)$, is much larger than the shifts obtained from the unwanted couplings. For instance, for the $\mathcal{H}_{sp}$ subspace, the following condition must hold:
\begin{equation}
\frac{ \left\vert \left\langle ns;np \left\vert V_{dd} \right\vert n_1 p; n_2 d \right\rangle \right\vert ^2}{\left\vert  E^{(0)}_{(ns;np)} - E^{(0)}_{(n_1 p;n_2d)} \right\vert } \ll min\left( \Delta^{(1)}_{sp} \right) \label{neglcond}. 
\end{equation}
\newline \indent Let us now see how the previous situation can be implemented in a rubidium atom system. Assuming $n=42$ and $\overrightarrow
{R}_{12}=\overrightarrow
{R}_{23}=R\overrightarrow{e}_{z}$ with $R=5 \mu m$, we numerically checked both the non-resonance and the negligibility conditions Eq.(\ref{neglcond}) for the unwanted couplings listed in Table \ref{TabRub}. Then we calculated the dipole-dipole interaction induced shifts in the three degenerate subspaces $\mathcal{H}_{sp}$, $\mathcal{H}_{pd}$ and $\mathcal{H}_{spd}$, using $\mathcal{R}_{ns}^{np} \simeq 2645 a_{0}$ and $\mathcal{R}_{np}^{nd} \simeq 2644 a_{0}$ where $a_{0}$ is the Bohr radius $\left(
a_{0}\simeq5.3\times10^{-11}m\right)  $, which yielded
\ \begin{table}[ptb]
\caption{Relevant states and unwanted couplings.}%
\label{TabRub}
\begin{indent}
\begin{tabular}{|l|}
\hline
\hspace{1.6cm}unwanted couplings \\
\hline
$   \left \vert ns,np \right \rangle \leftrightarrow  \left \vert n_{1}p,n_{2}s \right \rangle$, for $(n_1,n_2)\neq(n,n)$ \\
$   \left \vert ns,np \right \rangle \leftrightarrow  \left \vert n_{1}p,n_{2}d \right \rangle$ \\
$   \left \vert ns,nd \right \rangle \leftrightarrow  \left \vert n_{1}p,n_{2}p \right \rangle$ \\
$   \left \vert ns,nd \right \rangle \leftrightarrow  \left \vert n_{1}p,n_{2}f \right \rangle$ \\
$   \left \vert np,nd \right \rangle \leftrightarrow  \left \vert n_{1}s,n_{2}p \right \rangle$ \\
$   \left \vert np,nd \right \rangle \leftrightarrow  \left \vert n_{1}s,n_{2}f \right \rangle$ \\
$   \left \vert np,nd \right \rangle \leftrightarrow  \left \vert n_{1}d,n_{2}p \right \rangle$, for $(n_1,n_2)\neq(n,n)$ \\
$   \left \vert np,nd \right \rangle \leftrightarrow  \left \vert n_{1}d,n_{2}f \right \rangle$ \\
\hline
\end{tabular}
\end{indent}
\end{table}
$\left\vert \Delta_{sp} \right \vert = \simeq 6.1 \times 10^{-4}
- 1.2 \times 10^{-3} cm^{-1} \simeq 18 - 36 MHz$,
$\left\vert\Delta_{pd} \right\vert \simeq 4.2\times 10^{-5}- 1.2 \times 10^{-3} cm^{-1} \simeq 1.3 - 35 MHz$, and
$\left\vert\Delta_{spd}\right\vert \simeq 6.7\times 10^{-5}- 2.2 \times 10^{-3} cm^{-1} \simeq 2 - 67 MHz$. If the Rabi frequencies of the laser beams remain small compared to these shifts (here, typically, $1MHz$) the excitation of the corresponding states will be blocked. Conversely it puts a lower bound of the order of $1 \mu s$ for the typical time duration of single atom operations.  
\section{Conditional dynamics in a system of individual atoms}
\begin{figure}
[ptb]
\begin{center}
\includegraphics[
height=2in,
width=2.3in
]%
{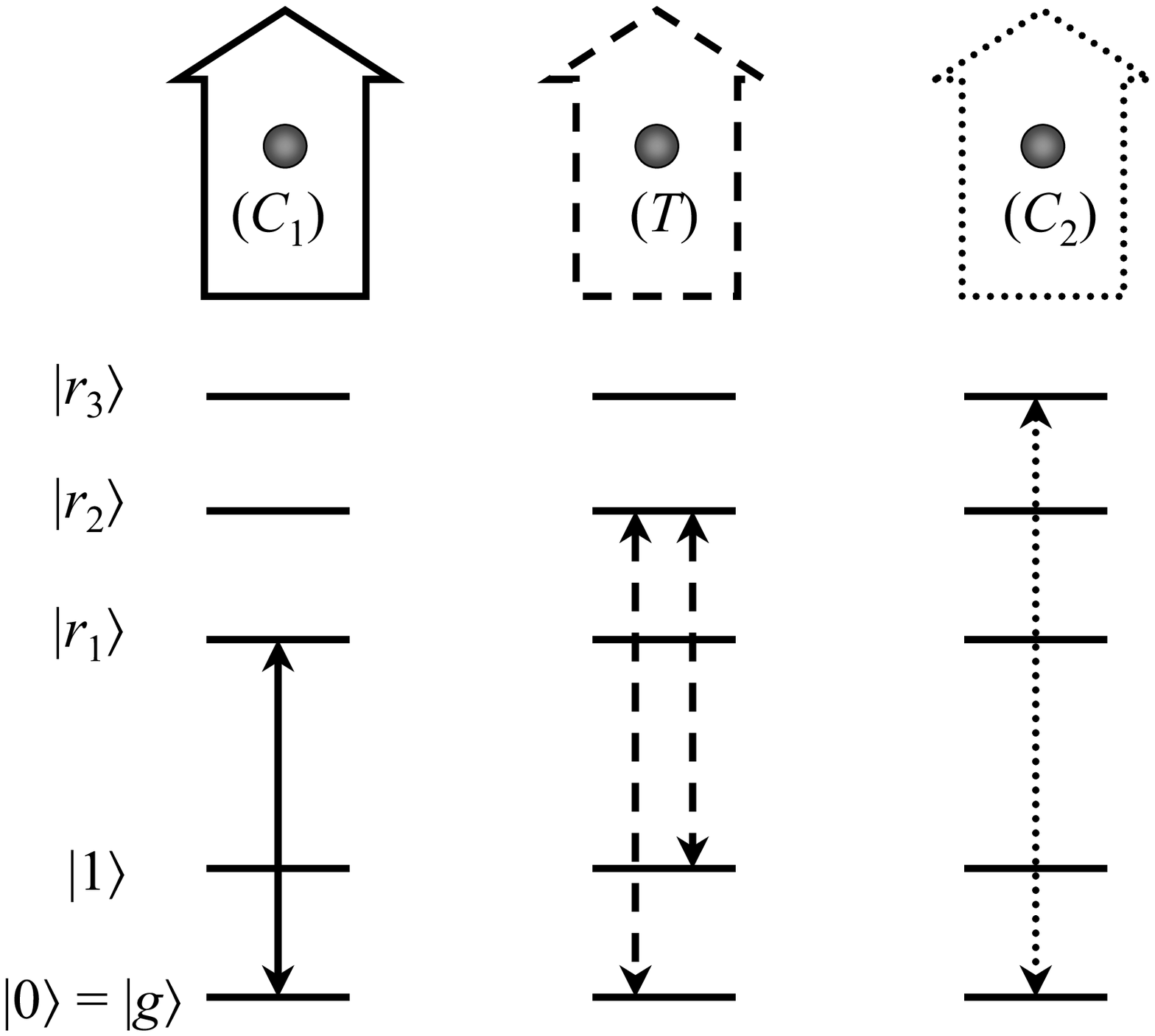}%
\caption{Level scheme and laser excitations for implementing the \textsc{Toffoli} gate in the three-atom system.}
\label{fig2}%
\end{center}
\end{figure}

\begin{figure*}
[ptb]
\begin{center}
\includegraphics[
height=2.8712in,
width=4.3076in
]%
{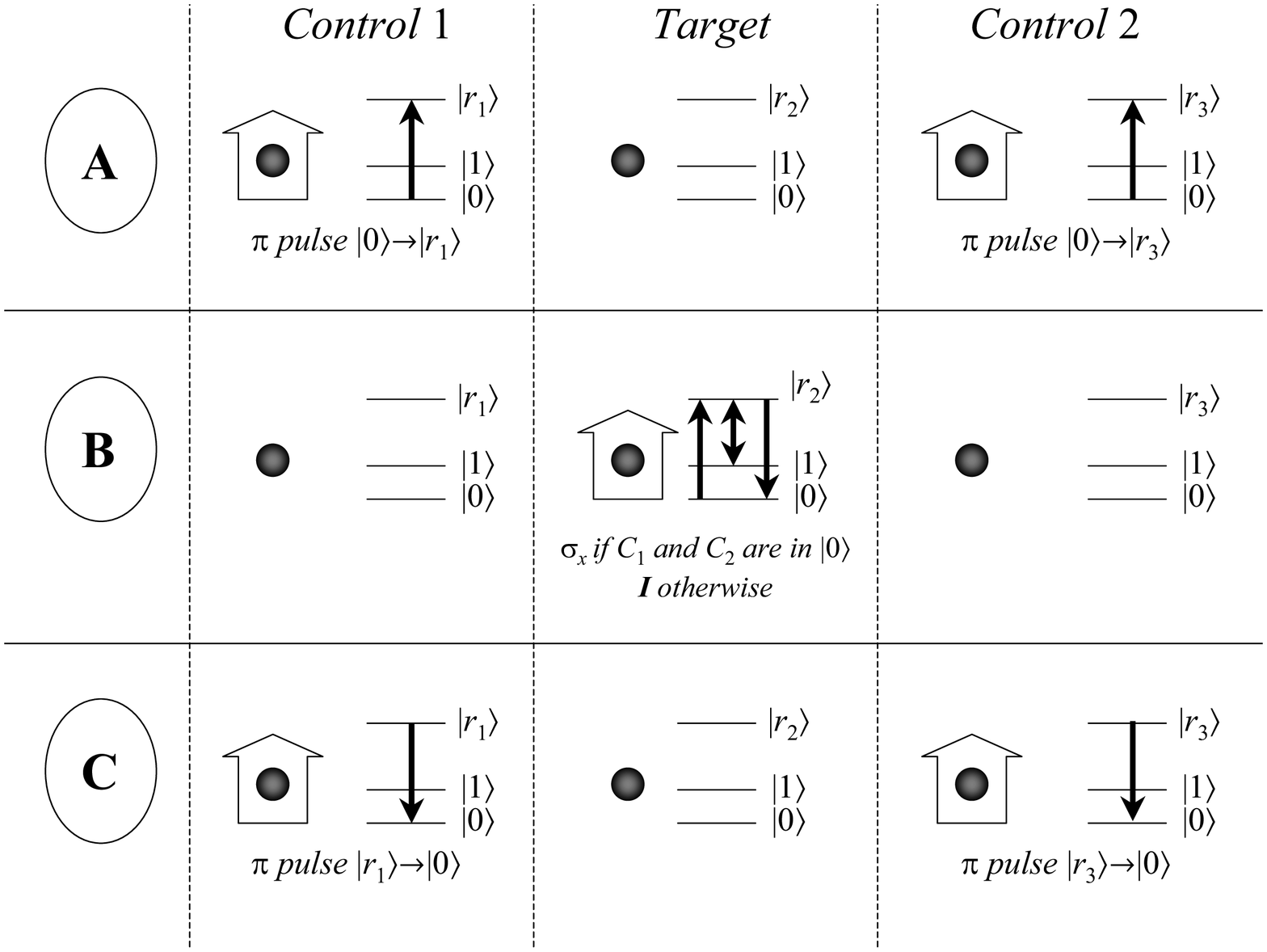}%
\caption{Three-step implementation of a \textsc{Toffoli} gate in a three-atom
system involving 3 different Rydberg states.}%
\label{fig3}%
\end{center}
\end{figure*}

\indent Let us now see how to use the spectroscopic situation described in the previous section in order to induce conditional dynamics in individual atom systems. To be specific, here, we show how to implement a three-qubit \textsc{Toffoli} gate \cite{NC00} in the three-atom system considered above. A qubit of information is encoded in each of the three atoms on the ground state $\left\vert 0\right\rangle \equiv\left\vert g\right\rangle $\ and a
low-lying excited state $\left\vert 1\right\rangle \equiv\left\vert
q\right\rangle $, the Rydberg states will be only temporarily populated during the gate, to achieve conditional dynamics through dipole-dipole interaction induced shifts (see Fig.\ref{fig2}). Atoms $(1), (2), (3)$ will respectively play the roles of Control 1-  $\left(  C_{1}\right)  $, Target- $\left(  T\right)
$ and Control 2-qubits $\left(  C_{2}\right)$. 
\newline \indent The \textsc{Toffoli} gate is then implemented through the following three-step procedure (see
Fig. \ref{fig3}). \newline \indent A. One first submits control atoms $\left(  C_{1}\right)  $
and\ $\left(  C_{2}\right)  $ to $\pi$-laser pulses which couple $\left\vert
0\right\rangle $ to $\left\vert r_{1}\right\rangle $ and $\left\vert
r_{3}\right\rangle $, respectively. \newline \indent B. One successively applies three $\pi
$-laser pulses on the target atom $\left(  T\right)  $ which couple
$\left\vert 0\right\rangle $ to $\left\vert r_{2}\right\rangle $, $\left\vert
1\right\rangle $ to $\left\vert r_{2}\right\rangle $ and $\left\vert
0\right\rangle $ to $\left\vert r_{2}\right\rangle $, respectively (in the
absence of the control atoms, this boils down to performing the Pauli matrix
$\sigma_{x}$ in the computational basis $\left\vert 0\right\rangle ,\left\vert
1\right\rangle $). \newline \indent C. One repeats the first step.
\newline If (at least) one of the atoms $\left(  C_{1}\right)  $ and\ $\left(
C_{2}\right)  $ is initially in the state $\left\vert 0\right\rangle $, at the
end of step A, (at least) one of the states $\left\vert r_{1}\right\rangle $
and $\left\vert r_{3}\right\rangle $ is excited: the shift induced by the
dipole-dipole interaction between $\left(  T\right)  $ and $\left(
C_{1}\right)  $ and/or\ $\left(  C_{2}\right)  $ will then prevent the target
atom from being excited into the state $\left\vert r_{2}\right\rangle $,
\textit{i.e.} step B will not induce any change. In contrast, if both
$\left(  C_{1}\right)  $ and\ $\left(  C_{2}\right)  $ are initially in
$\left\vert 1\right\rangle $, no interaction will shift the state $\left\vert
r_{2}\right\rangle $, step B will thus result in a $\sigma_{x}$ gate on the
Target atom. Finally, the overall transformation is thus a \textsc{Toffoli}
gate
\[
\text{\textsc{Toffoli}}=\text{\textsc{ccnot}}=\left(
\begin{array}
[c]{cc}%
\mathbb{I}_{6} & 0\\
0 & \sigma_{x}
\end{array}
\right)
\]
expressed in the computational basis $\left\vert c_{1}c_{2}t\right\rangle
=\left\vert 000\right\rangle ,\left\vert 001\right\rangle ,\left\vert
010\right\rangle ,\left\vert 011\right\rangle ,\left\vert 100\right\rangle
,\left\vert 101\right\rangle ,\left\vert 110\right\rangle ,\left\vert
111\right\rangle $. Note that, even though the same result can be achieved through combining one- and two- qubit elementary gates, our proposal only involves three steps and thus constitutes a more economical implementation of the \textsc{Toffoli} gate. 
\section{Conditional dynamics in
mesoscopic ensembles} 
\begin{figure}
[ptb]
\begin{center}
\includegraphics[
height=1.6in,
width=2.5in
]%
{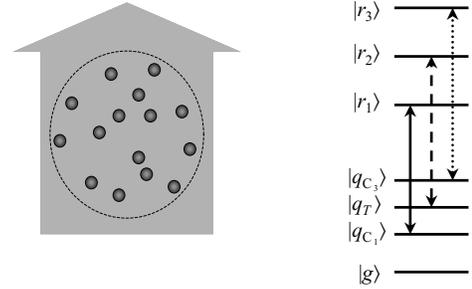}%
\caption{Level scheme and laser excitations for implementing the \textsc{ccphase} gate in an atomic ensemble.}
\label{fig4}%
\end{center}
\end{figure}

\indent The same kind of conditional dynamics can also
be performed on qubits stored in mesoscopic ensembles. To be specific, here, we shall show how to perform a \textsc{ccphase} gate in an ensemble made of the same atoms as in Section \ref{sec2}. In addition to the ground state $\left\vert g \right\rangle$ and Rydberg states $\left \vert r_1 \right \rangle$, $\left \vert r_2 \right \rangle$, and $\left \vert r_3 \right \rangle$, we shall need three extra long-lived atomic states $\left \vert q_{C_1} \right \rangle$, $\left \vert q_{T} \right \rangle$, and $\left \vert q_{C_2} \right \rangle$ (see Fig.\ref{fig4}). Following \cite{Klaus}, we encode three qubits of information on the eight collective states $
\left\vert 000\right\rangle \equiv \left\vert \mathbf{g}\right\rangle$, $
\left\vert 100\right\rangle  \equiv \left\vert \mathbf{q}_{C_{1}}^{1}\right\rangle$, $
\left\vert
010\right\rangle  \equiv \left\vert \mathbf{q}_{C_{2}}^{1}\right\rangle$, $
\left\vert
001\right\rangle  \equiv  \left\vert \mathbf{q}_{T}^{1}\right\rangle$, $
\left\vert
110\right\rangle  \equiv \left\vert \mathbf{q}_{C_{1}}^{1}\mathbf{q}_{C_{2}}^{1}\right\rangle$, $
\left\vert 101\right\rangle  \equiv \left\vert
\mathbf{q}_{C_{1}}^{1}\mathbf{q}_{T}^{1}\right\rangle$, $
\left\vert
011\right\rangle  \equiv \left\vert \mathbf{q}_{C_{2}}^{1}\mathbf{q}_{T}^{1}
\right\rangle$, and $ 
\left\vert 111\right\rangle  \equiv \left\vert
\mathbf{q}_{C_{1}}^{1}\mathbf{q}_{C_{2}}^{1}\mathbf{q}_{T}^{1}\right\rangle$
where $\left\vert \mathbf{q}_{C_{1}}^{1}\mathbf{q}_{C_{2}}%
^{1}\right\rangle $\ (for instance) denotes the symmetric collective
state with one atom in $\left\vert q_{C_{1}}\right\rangle $ and another in $\left\vert q_{C_{2}}\right\rangle $.
\indent The following three-step procedure implements the desired \textsc{ccphase} gate.
\newline \indent A. One first submits the whole sample to two $\pi$ laser pulses which couple
$\left\vert q_{C_{1}}\right\rangle $ and $\left\vert q_{C_{2}}\right\rangle $
to the Rydberg states $\left\vert r_{1}\right\rangle $ and $\left\vert
r_{3}\right\rangle $, respectively. This induces the following
transformations: $\left\vert \mathbf{g}\right\rangle \rightarrow\left\vert
\mathbf{g}\right\rangle $, $\left\vert \mathbf{q}_{C_{1}}^{1}\right\rangle
\rightarrow\left\vert \mathbf{r}_{1}^{1}\right\rangle $, $\left\vert
\mathbf{q}_{T}^{1}\right\rangle \rightarrow\left\vert \mathbf{q}_{T}%
^{1}\right\rangle $, $\left\vert \mathbf{q}_{C_{2}}^{1}\right\rangle
\rightarrow\left\vert \mathbf{r}_{3}^{1}\right\rangle $, $\left\vert
\mathbf{q}_{C_{1}}^{1}\mathbf{q}_{T}^{1}\right\rangle \rightarrow\left\vert
\mathbf{r}_{1}^{1}\mathbf{q}_{T}^{1}\right\rangle $, $\left\vert
\mathbf{q}_{C_{1}}^{1}\mathbf{q}_{C_{2}}^{1}\right\rangle \rightarrow
\left\vert \mathbf{r}_{1}^{1}\mathbf{r}_{3}^{1}\right\rangle $, $\left\vert
\mathbf{q}_{T}^{1}\mathbf{q}_{C_{2}}^{1}\right\rangle \rightarrow\left\vert
\mathbf{q}_{T}^{1}\mathbf{r}_{3}^{1}\right\rangle $, $\left\vert
\mathbf{q}_{C_{1}}^{1}\mathbf{q}_{C_{2}}^{1}\mathbf{q}_{T}^{1}\right\rangle
\rightarrow\left\vert \mathbf{r}_{1}^{1}\mathbf{r}_{3}^{1}\mathbf{q}_{T}%
^{1}\right\rangle$. \newline \indent B. One then applies a $2\pi$-pulse on the ensemble
which couple $\left\vert q_{T}\right\rangle $ to $\left\vert r_{2}%
\right\rangle $. The pulse will cause a transition in the ensemble only if none of the
Rydberg levels $\left\vert r_{1}\right\rangle $ and $\left\vert r_{3}%
\right\rangle $ are excited: it follows that only the state $\left\vert
\mathbf{q}_{b}^{1}\right\rangle$ will be multiplied by $-1$, the
other states being left unchanged. \newline \indent C. Finally, one applies the same two $\pi$
laser pulses as in the first step, which induces the inverse transformations.
The overall transformation is thus a \textsc{ccphase}, which imposes a
$\sigma_{z}$ gate on the target qubit initially stored in the $T$ ensemble
\textit{iff} the control ensembles $C_{1}$\ and $C_{2}$\ are initially in
state $\left\vert 0\right\rangle $. 
\section{Conclusion} In
this paper, we proposed new configurations for dipole-dipole Rydberg
interactions, involving different coupled and non-coupled Rydberg states. We think that such configurations are very promising and should allow for efficient implementation of sophisticated conditional dynamics beyond two-qubit gates. As first examples, we showed how to perform two specific three-qubit gates (the \textsc{ccnot} and \textsc{ccphase} gates) in an individual atom system and in an atomic ensemble, through appropriately exciting atoms into three different Rydberg states. The
feasibility of our schemes has been verified for the specific example of
rubidium atoms.\newline\indent We are currently investigating how such configurations with different Rydberg states could contribute, on the one hand, to extend the blockade phenomenon to macroscopic ensembles, and, on the other hand, to solve grid games as Latin squares quantum-mechanically.

\section*{Acknowledgements}

This work was supported by ARO-DTO grant nr. 47949PHQC and the European Union integrated project SCALA. The authors thank E. Bonderup, A.
Negretti and M. Saffman for fruitful discussions.

\appendix

\section{Calculation of the radial integral} \label{appR}
The radial part of the hydrogenic wavefunction takes the expression
\cite{Messiah} $R_{nl}\left(  r\right) =a^{-\frac{3}{2}}N_{nl}F_{nl}\left(  \frac{2r}%
{na}\right)$, with $a  =\frac{a_{0}}{Z}=\frac{\hbar^{2}}{Zm^{\prime}e^{2}}$,
$ N_{nl} =\frac{2}{n^{2}}\sqrt{\frac{\left(  n-l-1\right)  !}{\left[  \left(
n+l\right)  !\right]  ^{3}}}$ and $F_{nl}\left(  x\right) =x^{l}e^{-\frac{x}{2}}L_{n-l-1}^{2l+1}\left(
x\right)$, where $L_{p}^{k}\left(  x\right)$ is an associated Laguerre polynomial. One can thus put
the radial integral $\mathcal{R}_{n,l}^{n^{\prime},l^{\prime}} = \int_{0}^{+\infty}%
dr \ r^{3}R_{nl}\left(  r\right)  R_{n^{\prime}l^{\prime}}\left(  r\right)$ in the form
\begin{eqnarray*}
& \mathcal{R}_{n,l}^{n^{\prime},l^{\prime}}=a\times\frac{2^{l+l^{\prime}}N_{nl}
N_{n^{\prime}l^{\prime}}n^{4+l^{\prime}}\left(  n^{\prime}\right)  ^{4+l}
}{\left(  n+n^{\prime}\right)  ^{4+l+l^{\prime}}} 
\\
& \times \int_{0}^{+\infty
}dx \ x^{3+l+l^{\prime}}e^{-x} \\ 
& \times L_{n-l-1}^{2l+1}\left(  \frac{2x}{1+n/n^{\prime}%
}\right)  L_{n^{\prime}-l^{\prime}-1}^{2l^{\prime}+1}\left(  \frac
{2x}{1+n^{\prime}/n}\right) 
\end{eqnarray*}
Using the definition $L_{p}^{k}\left(  x\right)  =\sum_{s=0}^{p}\left(
-1\right)  ^{s}\frac{\left[  \left(  p+k\right)  !\right]  ^{2}}{\left(
p-s\right)  !\left(  k+s\right)  !s!}x^{s}$, one finally gets the explicit
expression
\begin{widetext}
\begin{align*}
\mathcal{R}_{n,l}^{n^{\prime},l^{\prime}}  & =a\times2^{l+l^{\prime}+2}\frac
{n^{2+l^{\prime}}\left(  n^{\prime}\right)  ^{2+l}}{\left(  n+n^{\prime
}\right)  ^{4+l+l^{\prime}}}\sqrt{\left(  n+l\right)  !\left(  n^{\prime
}+l^{\prime}\right)  !\left(  n-l-1\right)  !\left(  n^{\prime}-l^{\prime
}-1\right)  !}\\
& \times\sum_{r=0}^{n-l-1}\sum_{s=0}^{n^{\prime}-l^{\prime}-1}\left(
-\frac{2}{n+n^{\prime}}\right)  ^{r+s}\frac{n^{s}\left(  n^{\prime}\right)
^{r}}{r!s!}\frac{\left(  3+l+l^{\prime}+r+s\right)  !}{\left(  n-l-r-1\right)
!\left(  2l+r+1\right)  !\left(  n^{\prime}-l^{\prime}-s-1\right)  !\left(
2l^{\prime}+s+1\right)  !}.
\end{align*}
\end{widetext}

\end{document}